\documentclass[12pt]{article}

\usepackage[numbers,sort&compress]{natbib}
\usepackage{graphicx}
\usepackage{amsmath}
\usepackage{amssymb}
\usepackage{hyperref}

\ifx\hypersetup\sadfkjashdfkxja\else
\hypersetup{pdftitle={Wilson loops in Five-Dimensional Super-Yang-Mills}}
\hypersetup{pdfauthor={Donovan Young}}
\hypersetup{pdfsubject={}}
\hypersetup{pdfkeywords={}}
\hypersetup{plainpages=false}
\hypersetup{bookmarksnumbered=true}
\hypersetup{pdfstartview=FitH}
\hypersetup{pdfpagemode=UseNone}
\hypersetup{colorlinks=false}
\hypersetup{citebordercolor={.5 1 .5}}
\hypersetup{urlbordercolor={.5 1 1}}
\hypersetup{linkbordercolor={1 .7 .7}}
\fi

\def\be{\begin{equation}}
\def\ee{\end{equation}}
\def\bsp{\be\begin{split}}
\def\la{\langle}
\def\ra{\rangle}

\def\G{\Gamma}

\def\S{\Sigma}
\def\a{\alpha}
\def\b{\beta}
\def\g{\gamma}

\def\d{\delta}
\def\e{\epsilon}
\def\m{\mu}

\def\n{\nu}
\def\s{\sigma}

\def\t{\tau}
\def\o{\omega}
\def\O{\Omega}
\def\T{\theta}

\def\p{\partial}

\def\bR {\mathbb{R}}

\def\vp{\varphi}

\makeatletter

\newcommand{\Rmnum}[1]{\expandafter\@slowromancap\romannumeral #1@}
\makeatother

\newcommand{\beq}{\begin{equation}}
\newcommand{\eeq}{\end{equation}}
\newcommand{\bea}{\begin{eqnarray}}
\newcommand{\eea}{\end{eqnarray}}

\renewcommand{\title}[1]{\vbox{\center\LARGE{#1}}\vspace{5mm}}
\renewcommand{\author}[1]{\vbox{\center\large{#1}}\vspace{5mm}}
\newcommand{\address}[1]{\vbox{\center\em#1}}
\newcommand{\email}[1]{\vbox{\center\tt#1}\vspace{5mm}}

\newcommand{\Tr}{\mathrm{Tr}}

\newcommand{\Cset}{{\,\,{{{^{_{\pmb{\mid}}}}\kern-.47em{\mathrm C}}}}}

\newcommand{\comment}[1]{}
\begin{document}
\bibliographystyle{utphys}
\newpage
\setcounter{page}{1}
\pagenumbering{arabic}
\renewcommand{\thefootnote}{\arabic{footnote}}
\setcounter{footnote}{0}

\begin{titlepage}
\title{\vspace{1.0in} {\bf Wilson loops in Five-Dimensional Super-Yang-Mills}}
 
\author{Donovan Young}

\address{Niels Bohr Institute, University of Copenhagen, Blegdamsvej 17, DK-2100 Copenhagen, Denmark}

\email{dyoung@nbi.dk}

\abstract{We consider circular non-BPS Maldacena-Wilson loops in
  five-dimensional supersymmetric Yang-Mills theory ($d=5$ SYM) both
  as macroscopic strings in the D4-brane geometry and directly in
  gauge theory. We find that in the D$p$-brane geometries for
  increasing $p$, $p=4$ is the last value for which the radius of the
  string worldsheet describing the Wilson loop is independent of the
  UV cut-off. It is also the last value for which the area of the
  worldsheet can be (at least partially) regularized by the standard
  Legendre transformation. The asymptotics of the string worldsheet
  allow the remaining divergence in the regularized area to be
  determined, and it is found to be logarithmic in the UV cut-off.  We
  also consider the M2-brane in $AdS_7\times S^4$ which is the
  M-theory lift of the Wilson loop, and dual to a ``Wilson surface''
  in the $(2,0)$, $d=6$ CFT.  We find that it has exactly the same
  logarithmic divergence in its regularized action. The origin of the
  divergence has been previously understood in terms of a conformal
  anomaly for surface observables in the CFT. Turning to the gauge
  theory, a similar picture is found in $d=5$ SYM. The divergence and
  its coefficient can be recovered for general smooth loops by summing
  the leading divergences in the analytic continuation of dimensional
  regularization of planar rainbow/ladder diagrams. These diagrams are
  finite in $5-\e$ dimensions.  The interpretation is that the Wilson
  loop is renormalized by a factor containing this leading divergence
  of six-dimensional origin, and also subleading divergences, and that
  the remaining part of the Wilson loop expectation value is a finite,
  scheme-dependent quantity. We substantiate this claim by showing
  that the interacting diagrams at one loop are finite in our
  regularization scheme in $d=5$ dimensions, but not for $d\geq 6$.}

\end{titlepage}

\section{Introduction, main results, and conclusions}

The Maldacena-Wilson loop \cite{Maldacena:1998im,Rey:1998ik} continues
to be a remarkably useful observable in the context of the
gauge-gravity duality. In the context of AdS/CFT, the circular Wilson
loops
\cite{Erickson:2000af,Drukker:2000rr,Drukker:2008zx,Chen:2008bp,Rey:2008bh,Drukker:2009hy}
have proven to be amenable to exact calculation using the techniques
of localization
\cite{Pestun:2007rz,Kapustin:2009kz,Marino:2009jd,Drukker:2010nc},
providing hard predictions for a range of stringy and M-theoretic
phenomena including semi-classical fundamental strings and membranes,
D-branes, and bubbling geometries
\cite{Drukker:2005kx,Gomis:2006sb,Gomis:2006im,Hartnoll:2006is,Hartnoll:2006ib,Yamaguchi:2006tq,D'Hoker:2007fq,Yamaguchi:2006te,Lunin:2006xr}. Continuing
these successes outside the regime of conformal symmetry, in
particular to maximally supersymmetric Yang-Mills (SYM) theories in
dimensions other than four, is an important step towards understanding
the gauge-gravity duality in these far-less-explored contexts.

In this paper we will consider the circular Maldacena-Wilson loop with
constant scalar coupling (see (\ref{wl}) for a definition) in $d=5$
SYM. In this five-dimensional context the circular Wilson loop
preserves no global supersymmetries. The first and most obvious
question is whether the $d=5$ theory is sensible, since by standard
power-counting it is a non-renormalizable theory. We will follow the
procedure of using dimensional reduction from ${\cal N}=1$, $d=10$ SYM
to $2\o$ dimensions. Since we are {\it above} rather than below four
dimensions (where dimensional regularization actually renders
integrals over loop momenta UV-finite), this procedure is viewed as an
analytic continuation from convergent results at $d<4$ to
$d=5$. Despite this questionable regularization scheme, we find that
we can make contact with the string dual, i.e. a fundamental string in
the D4-brane geometry \cite{Itzhaki:1998dd}, and its M2-brane lift.

\begin{figure}\label{fig:taken}
\begin{center}
\includegraphics*[bb= 0 0 685 430,width=3.5in]{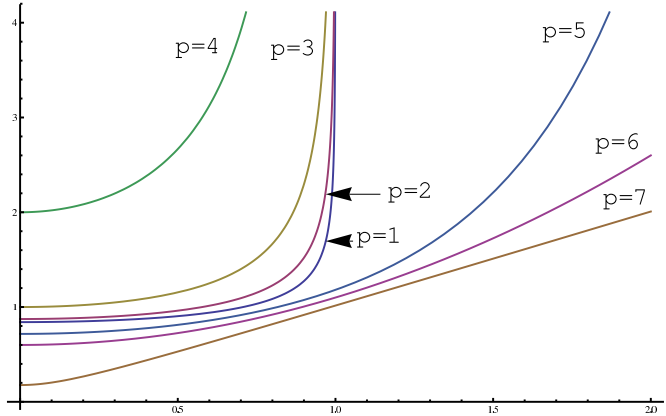}
\end{center}
\caption{Profiles of string duals of 1/4 BPS circular Wilson loops
  (\ref{quart}) in the D$p$-brane geometries (\ref{pbrane}). The
  holographic $U$ direction is the vertical axis, the boundary radial
  coordinate $r$ is the horizontal axis. There is a marked difference
  between $p>4$ and $p\leq 4$. Taken from \cite{Agarwal:2009up}.}
\end{figure}

It appears that in the D$p$-brane geometries, for increasing $p$, $p=4$ (corresponding to
$d=5$ SYM) is in some sense a final outpost. In the work
\cite{Agarwal:2009up}, the embedding functions for strings dual to 1/4
BPS circular Wilson loops with trivial expectation value (a
generalization of the Zarembo loops \cite{Zarembo:2002an}) in SYM for
$2 \leq d \leq 8$ were presented. There is a stark division in the
behaviour of the worldsheets as they approach the boundary of the
geometry precisely between $p=4$ and $p\geq 5$. Specifically, for
$p\leq 4$, the radius $R$ of the worldsheet (dual to the radius of the
Wilson loop contour), assumes a constant value as the boundary is
approached. For $p>4$ this is no longer true and a UV cut-off must
explicitly be added in order to define $R$ -- or equivalently -- $R$
becomes a function of the UV cut-off, see figure \ref{fig:taken},
taken from \cite{Agarwal:2009up}.
The figure shows a plot of the
string profile with the holographic direction plotted on the vertical
axis, and a boundary radial direction plotted on the horizontal
axis. In section \ref{bps} we review the details of these solutions.

It turns out that this behaviour is generic, in the sense that it persists
for the duals of non-BPS circular Wilson loops, which unlike the 1/4 BPS ones,
have a constant coupling to the scalar fields of SYM. In section
\ref{nonbps}, we carry out the analysis of the string worldsheets for
these Wilson loops. Although we are unable to find exact solutions
for the embeddings, we can find their asymptotic behaviour as the
boundary is approached. Using this information we show that $R$ is
independent of the UV cut-off only for $p<5$, as for the BPS
solutions.

The area of the worldsheets describing the Wilson loops is infinite
for any $p$. In the $AdS_5\times S^5$ case (i.e. $p=3$), this
divergence is well understood and is removed by a Legendre
transformation \cite{Drukker:1999zq}. In the case of the 1/4 BPS
loops in the D$p$-brane geometries, the same Legendre transformation
simply eliminates the area entirely, giving the trivial expectation
value for the Wilson loop $e^0=1$. For the non-BPS Wilson loops, we
find that again $p=4$ is a special value. It is the last value for
which the Legendre transformation removes the (in this case leading)
divergence of the area, leaving a $\log(\text{UV cut-off})$ divergence
and a finite piece. The asymptotics of the string worldsheet allow the
determination of the coefficient of the $\log$, while the finite piece
is not obtainable. It is, in any case not well defined, since a shift
in the UV cut-off would affect it. The details of these calculations
are given section \ref{nonbps}. We quote the result here for
convenience\footnote{Note that $U$ has dimensions of energy. There is
  of course a scale (not shown) giving a dimensionless argument for
  the logarithm. It is set by the minimum value of $U$ plumbed by the
  worldsheet, which is in turn related to the radius of the loop.}
\be
\la W \ra =(\text{prefactor})\cdot\exp\left(-S_{\text{reg.}}\right) 
=(\text{prefactor})\cdot\exp\left(\frac{g^2N}{16\pi R}\log
U_{\text{max.}}\right)\cdot(\text{finite}), 
\ee
where $U_{\text{max.}}$ is the cut-off in the holographic direction,
see (\ref{pbrane}), and where ``prefactor'' refers to the
semi-classical dressing of the main exponential portion of the
partition function, see \cite{Agarwal:2009up} for a discussion.

The fact that the $p=4$ case shares these two ``nice'' features with
its lower dimensional cousins, i.e. regularizable worldsheet area and
radius independent of the cut-off, certainly resonates with recent
speculations about the possible finiteness of $d=5$ SYM
\cite{Lambert:2010iw,Douglas:2010iu}. We will see that, using
dimensional regularization analytically continued to five-dimensions,
the gauge theory makes contact with this behaviour of the string
dual. Specifically, we perform an analysis of the planar
rainbow/ladder diagrams in section \ref{gauge}. The ``loop-to-loop''
propagator $P(\t_1,\t_2)$ (\ref{loop2loop}) has the following
behaviour
\be
P(\t_1,\t_2) \propto \frac{1}{\sin^{2\o-4} \frac{\t_{12}}{2}},
\ee
where $\t_{12}=\t_1-\t_2$ is the difference between the Wilson loop
contour parameter at the two points where the propagator is joined,
and the dimension $d=2\o$. We
find that a certain sub-class of planar rainbow/ladder diagrams
contributes the highest divergence, order-by-order in the perturbative
expansion. This allows us to sum-up all of these contributions,
finding
\be
\la W \ra_{\text{perturbative}} = (\text{prefactor})\cdot
\exp\left(\frac{g^2N}{16\pi R} \frac{1}{\e}\right)\cdot(\text{finite}),
\ee
where $\e = 5-2\o$. Equating $\log U_{\text{max.}}$ with $1/\e$ we
find an exact match. In fact if we set $2\o = 5$ and use instead a
point-splitting regularization $\d$ of the Wilson loop contour, we
obtain $-\log\d$ in place of $1/\e$, see section \ref{gauge}. We will
suggest that the subleading divergences should be associated with the
prefactor of the exponential term in the Wilson loop VEV.

Note that for $d=2\o$ slightly below $5$, the loop-to-loop propagator
integrates to a finite quantity. This appears to be a gauge theory
reflection of the fact that $d=5$ is the last sensible setting for the
Wilson loop in the string dual described above.  We might then expect
that the interacting diagrams are at least subleading, if not
finite. Evaluating the ${\cal O}(g^4)$ corrections, in section
\ref{int}, we find that they are indeed finite when $2\o$ is set
(i.e. analytically continued) to $5$, but not for $2\o \geq 6$. Again,
we see that the gauge theory picture is sensible (at least in our
regularization scheme), like in the string case, for the last time at
$d=5$, as the dimension is increased.

The D$4$-brane geometry cannot be trusted at arbitrarily close
distances to the boundary, and at a certain point the M-theory
description takes over\footnote{This happens for $g^2 U \sim
  N^{1/3}$.} and the background geometry becomes $AdS_7\times
S^4$. The lift of the non-BPS circular Wilson loop is an M2-brane and
is considered in section \ref{m2surf}.  The action of this M2-brane
may be regularized via an analogous Legendre transformation and has
precisely the same logarithmic term as the string in the D4-brane
geometry. This logarithmic divergence may be viewed as a conformal
anomaly of the dual surface operator in the $d=6$ $(2,0)$ CFT
\cite{Graham:1999pm,Henningson:1999xi,Gustavsson:2003hn,Gustavsson:2004gj}.
In section \ref{stringgen} we show that the regularized area of the
string worldsheet in the D$4$-brane geometry reproduces this anomaly
term for general-shaped smooth and closed contours; in section
\ref{shape} the corresponding generalization in the gauge theory is
performed.  It is remarkable that we have recovered this anomaly
directly in $d=5$ SYM, especially given recent speculations that the
$d=6$ CFT might be captured entirely by $d=5$ SYM
\cite{Lambert:2010iw,Bolognesi:2011nh,Bolognesi:2011rq,Douglas:2010iu,Gustavsson:2011ur}.

There are various extensions of the present work which could be
considered. There is a kind of conformal symmetry at play in the
D$p$-brane backgrounds and the associated SYM theories
\cite{Jevicki:1998ub}. It would be interesting to understand whether
this symmetry can be used to obtain the circular Wilson loop
considered here as a transformation of the 1/2 BPS straight line
\cite{Agarwal:2009up} (which has trivial expectation value), as is the
case in $d=4$ \cite{Drukker:2000rr}. This may provide an alternate
derivation of the leading divergence in terms of a (generalized)
conformal anomaly (in this case directly in $d=5$ SYM).  The
exponential factor dressing the Wilson loop expectation value is very
reminiscent of Wilson loop renormalization in four dimensions
\cite{Frenkel:1984pz,Gatheral:1983cz,Brandt:1981kf,Polyakov:1980ca}. This
suggests that there is perhaps a way to give physical meaning to the
finite part of the expectation value, through a subtraction scheme. It
would also be interesting to consider correlators with local
operators. This calculation has been carried out for the spherical
Wilson surface in $AdS_7\times S^4$ in \cite{Corrado:1999pi}, where,
unlike for the expectation value of the surface itself, finite results
are obtained. Here one would require the explicit solution for the
M2-brane describing the Wilson loop; the techniques for computing
holographic correlation functions in the D$p$-brane backgrounds are
also available \cite{Kanitscheider:2008kd,Wiseman:2008qa}.  It is also
interesting to ask to what extent contact can be made with the 1/2 BPS
circular Wilson loop in ${\cal N}=4$, $d=4$ SYM through dimensional
reduction, especially as regards the S-duality of the latter as
discussed in \cite{Douglas:2010iu}, and whether localization
techniques have any application to the calculation of the
five-dimensional Wilson loop expectation value. Finally, we note that
we have not considered non-perturbative corrections to the Wilson
loop. It would be very interesting to explore their effect on the
expectation value.

\section{String duals of BPS and non-BPS Wilson loops}

In this section we will look at the generalization of the circular
Zarembo loops studied in \cite{Agarwal:2009up}, remarking that the
string duals require a cut-off in order to be defined in the
D$p$-brane backgrounds for $p>4$. We will then continue with a
generalization of these arguments to regular circular Wilson loops,
which, apart from the conformal case $p=3$, are non-BPS.  

\subsection{1/4 BPS Wilson loops}
\label{bps}

In the work \cite{Agarwal:2009up}, string solutions were found which
are dual to Maldacena-Wilson loops in $d$-dimensional SYM with
circular contours and with scalar couplings which also describe a
(great) circle
\be
x^\m=R\,(\cos\t,\,\sin\t,\,0,\ldots,0),\qquad
\Theta^I=(-\sin\t,\,\cos\t,\,0,\ldots,0).
\ee
The string duals are fundamental strings in the D$p$-brane geometries
\cite{Itzhaki:1998dd}\footnote{The definition of $C_p$ and the dilaton
  are consistent with the normalization of the SYM action
  $S=\frac{1}{g^2}\int d^{p+1}x \frac{1}{4} F_{\m\n}^a F^{a\m\n}+\ldots$.}
\bsp\label{pbrane}
&ds^2 = \a' \left( \frac{U^{(7-p)/2}}{C_p} dx_\shortparallel^2
+ \frac{C_p}{U^{(7-p)/2}} \, dU^2 + C_p\, U^{(p-3)/2}\,d\O_{8-p}^2
 \right),\\
 &e^\phi = (2\pi)^{1-p} g^2
 \left(\frac{C_p^2}{U^{7-p}}\right)^{(3-p)/4},~
 C_p^2 = g^2 N \,2^{6-2p} \pi^{(9-3p)/2}\, \G\left(\frac{7-p}{2} \right).
\end{split}
\ee
We let $dx_\shortparallel^2 = dr^2 + r^2d\vp^2 + dx_{\bR^{p-1}}^2$,
and $d\O_{8-p} = d\T^2 + \cos^2\T\,d\phi^2 + \sin^2\T \, d\O_{6-p}^2.$
We then set $\vp=\phi$ to one of our string worldsheet coordinates,
and let $r$, $U$, and $\theta$ depend only on the other one. 
Then the string solutions are expressed as (for $p\leq 7$)
\bsp\label{quart}
r(U) =
\begin{cases}
\sqrt{\frac{2\,C_p^2}{5-p}}\sqrt{ U_{\text{min.}}^{p-5}-U^{p-5}},~ &p < 5\\
\sqrt{\frac{2\,C_p^2}{p-5}}\sqrt{ U^{p-5}-U_{\text{min.}}^{p-5}},~ &p > 5\\
\sqrt{2 \,C_5^2 \,\log \frac{U}{U_{\text{min.}}}},~&p=5\\
\end{cases},\qquad
\sin\T = \frac{U_{\text{min.}}}{U},
\end{split}
\ee
and describe a string which wraps a hemisphere in the $S^{8-p}$ and
ends along a circular contour of radius $R$ at the boundary
($U=U_{\text{max.}}\gg 1$) of the remainder of the geometry. The
worldsheet smoothly closes-off at $U=U_{\text{min.}}$ where $r(U)=0$.

We now remark that in the cases where $p\geq 5$, the radius of the
Wilson loop depends upon the cut-off $U_{\text{max.}}$
\be
R=
\begin{cases}
\sqrt{\frac{2\,C_p^2}{5-p}}\sqrt{ U_{\text{min.}}^{p-5}},~ &p < 5\\
\sqrt{\frac{2\,C_p^2}{p-5}}\sqrt{ U_{\text{max.}}^{p-5}-U_{\text{min.}}^{p-5}},~ &p > 5\\
\sqrt{2 \,C_5^2 \,\log \frac{U_{\text{max.}}}{U_{\text{min.}}}},~&p=5\\
\end{cases}.
\ee
Thus the Wilson loops may be defined in the gravity duals to SYM in
$d\leq 5$ dimensions without recourse to a UV cut-off; the same is not
true for $d>5$.

The 1/4 BPS Wilson loops considered in this section are rather special
objects which have trivial (unit) expectation value\footnote{This is
  because the Legendre transformation (\ref{legendre}) exactly cancels the
  bare action \cite{Agarwal:2009up}.}. In 
section \ref{nonbps} we will consider non-BPS circles for which this is not the
case.

\subsection{Non-BPS Wilson loops}
\label{nonbps}

In section \ref{bps} we showed that certain 1/4 BPS circular Wilson
loops in $d$-dimensional SYM had string duals which could be defined
without recourse to a UV cut-off if $d=p+1\leq 5$. Here we will
consider a circular Wilson loop with a constant coupling to the
scalars, so that the string dual sits at a point on the $S^{8-p}$. For
$p\neq 3$ this object preserves no global supersymmetries. We will not
be able to solve for the string embedding, but we will derive its
asymptotic form as the boundary is approached. This we will use to
analyze the leading divergences in the worldsheet area.

We consider a fundamental string ending in a circular contour on the
boundary of the D$p$-brane geometry (\ref{pbrane}). We let
$dx_\shortparallel^2 = dr^2 + r^2d\vp^2 + dx_{\bR^{p-1}}^2$, and take
our string worldsheet to be parameterized by $U$ and $\vp$. Using the
ansatz whereby $r(U)$, the Nambu-Goto action is 
\be\label{strgact}
S = \int dU \,r\,\sqrt{1+\frac{U^{7-p}}{C_p^2}r'^2},
\ee 
where $f'\equiv\p_Uf$ and we have integrated over $\vp$ since the Lagrangian is
independent of it. One can then easily verify that for $p\leq 4$, near
the boundary at $U=\infty$
\be\label{psol}
r(U) = R - \frac{1}{5-p}\frac{C_p^2}{R}\frac{1}{U^{5-p}}+\ldots 
\ee
is a solution to the equation of motion. In the full solution, the
worldsheet closes-off at some minimum value of $U$,
$U_{\text{min.}}$. The radius $R$ is then related to
$U_{\text{min.}}$, in much the same way as the solutions presented in
section \ref{bps}. As we saw in that section, here there is a similar
marked difference for $p>4$. When we take $p>4$ we find that $r(U)$
diverges as $U\to\infty$, requiring the radius of the Wilson loop to
be defined at some cut-off $U_{\text{max.}}$. For example for $p=5$
one finds\footnote{Similar behaviour, i.e. $r(U)$ diverging as
  $U\to\infty$ is found for $p>5$.}
\bsp\label{p5}
&r(U) = R - \frac{C_5^2}{R}\log\frac{U_{\text{max.}}}{U} +\ldots.\\
\end{split}
\ee 
In this situation, as in section \ref{bps}, $R$ is a function of {\it
  both} $U_{\text{min.}}$ {\it and} $U_{\text{max.}}$. Thus the radius
of the Wilson loop is affected by changes to the cut-off. Given the
apparent non-renormalizability of supersymmetric Yang-Mills in $d>4$
dimensions, one would have expected this behaviour to set-in already
at $p=4$. The fact that it is postponed to $p\geq 5$ is interesting,
given recent speculations that SYM in $d=5$ may be a finite theory.

\subsubsection{Regularized area of the worldsheet}

We would now like to analyze the divergence in the area of the
worldsheet corresponding to the non-BPS circular Wilson loops. The
area of a Wilson loop is regularized via a Legendre transformation
using the ${\cal Y}$ coordinates defined as
\be \label{Ycoords}
\frac{dU^2}{U^2}+d\O_{8-p}^2 = \frac{d{\cal Y}^I d{\cal
    Y}^I}{{\cal Y}^2},\quad {\cal Y}^I = U \hat \T^I, \quad \hat \T^I
\hat \T^I=1, \quad I=1,\ldots,9-p.  \ee
Then the regularized area of the worldsheet $\S$ is defined as \cite{Drukker:1999zq}
\bsp\label{legendre}
S_{\text{reg.}} &= S - \int d\t\, d\s \,
\p_\s \left({\cal Y}^I \frac{\d {\cal L}}{\d \p_\s {\cal Y}^I} \right)\\
& = S - \int d\t\, {\cal Y}^I \frac{\d {\cal L}}{\d \p_\s {\cal Y}^I} \Biggr|_{\p\S},
\end{split}
\ee
where the $\t$ coordinate parametrizes the boundary contour, and
${\cal L}$ indicates the Lagrangian density, so that $S=\int d\t d\s
{\cal L}$. We then find that
\be
S_{\text{reg.}} = \int_{U_{\text{min.}}}^{U_{\text{max.}}} dU \,r\,\sqrt{1+\frac{U^{7-p}}{C_p^2}r'^2} - 
\frac{U\,r}{\sqrt{1+\frac{U^{7-p}}{C_p^2}r'^2}}\Biggr|_{U=U_{\text{max.}}}.
\ee
Specializing to $p=4$ we may calculate the regularized area of the
worldsheet using (\ref{psol}). One finds\footnote{The boundary term in
  the Legendre transformation also contributes to the finite piece.}
\bsp\label{p4reg}
&S = \int^{U_{\text{max.}}} dU \biggl(R -\frac{C_4^2}{2RU}+\ldots\biggr) = R\,
U_{\text{max.}} - \frac{C_4^2}{2R}\log
U_{\text{max.}}+\text{finite},\\
&S_{\text{reg.}} =  -\frac{C_4^2}{2R}\log
U_{\text{max.}}+\text{finite}.
\end{split}
\ee

The situation is vastly different for $p>4$. For example, for $p=5$,
using (\ref{p5}), we find that\footnote{The conditions under which a
  Legendre transform can remove the leading divergence has been analyzed
  in great detail in \cite{Chu:2008xg}. The results given here are a
  special case of that analysis.}
\be
S = \sqrt{C_5^2+R^2} \,U_{\text{max.}}+\text{finite},\qquad
\frac{U\,r}{\sqrt{1+\frac{U^{7-p}}{C_p^2}r'^2}}\Biggr|_{U=U_{\text{max.}}} = 
\frac{R^2}{\sqrt{C_5^2+R^2}}\,  U_{\text{max.}},
\ee
and therefore the regularization procedure does not remove the leading
divergence\footnote{Again, the same general behaviour, i.e. leading
  divergences not removed by the Legendre transformation, is found for $p>5$.}. This may be an indication that SYM
in dimensions only greater than five are non-renormalizable. 
Using (\ref{p4reg}), the Wilson loop expectation
value for $p=4$ is given by 
\be\label{string}
\la W_{\text{circle}} \ra = {\cal V} \,e^{-S_{\text{reg.}}}=
{\cal V} \exp\Biggl(\frac{g^2N}{16\pi R} \log
U_{\text{max.}}\Biggr) \cdot (\text{finite}),
\ee
where we have indicated the appearance of an unknown, and we will
argue from gauge theory, 
also divergent prefactor ${\cal V}$, which can in principle be determined using
semi-classical methods. 

In section \ref{gauge} we will recover the exponential factor in
(\ref{string}) by summing planar rainbow/ladder diagrams, and posit
that the remaining finite factor is provided by interacting diagrams.
Before doing so we would like to consider the uplift of the $p=4$ case
to M-theory, where we will see a six-dimensional origin of the
exponential factor. 

\subsubsection{M-theory lift}
\label{m2surf}

For strong coupling, defined as $g^2U\gg N^{1/3}$, the IIA D4-brane
geometry is replaced by the M-theory background $AdS_7\times S^4$ with
a boundary direction $x_6$ periodicially identified on a circle of
radius $R_6 = g^2/(8\pi^2)$ \cite{Itzhaki:1998dd}. The metric on this
space may be expressed as\footnote{\label{foot1}The coordinate $\tilde
  U$ is related to the $U$ coordinate of the D4-brane geometry via
  $\tilde U^2 = 2\pi U/(g^2 N)$, see \cite{Itzhaki:1998dd}.}
\be
ds^2 = 4(\pi N)^{2/3} l_p^2\,\Bigl(\frac{d\tilde U^2}{\tilde U^2} + \tilde U^2 \left(dr^2
+r^2d\phi^2+dx_6^2+dx_a^2\right) + \frac{1}{4} d\O_4^2
\Bigr),
\ee
where $a=1,\ldots,3$. We then consider an M2-brane with worldvolume
coordinates $\{\tilde U,\phi,x_6\}$ (i.e. wrapped on $x_6$) and take
$r(\tilde U)$. Shrinking the M-theory circle $x_6$ to
zero size, we recover the IIA fundamental string describing the
circular Wilson loop in the D4-brane geometry. The boundary surface of
the M2-brane
is $S^1\times S^1$, i.e. the Wilson loop circle times the $x_6$ circle. 
Integrating over $x_6$ and $\phi$ we obtain
\be\label{membact}
S_{M2} = 8 \pi N R_6 \int d\tilde U \, r\,\tilde U\,\sqrt{1+\tilde U^4 r'^2},
\ee
where we have used the M2-brane tension $T=l_p^{-3}/(2\pi)^2$. We then
find that the equation of motion for large $\tilde U$ is solved by
\be
r(\tilde U) = R - \frac{1}{4\tilde U^2 R} +\ldots.
\ee
The action of the M2-brane then evaluates to
\be\label{M2act}
\frac{S_{M2}}{8\pi N R_6} =\int^{\tilde U_{\text{max.}}} d\tilde U\, \left(R\,\tilde U -
\frac{1}{8\,R\,\tilde U}+\ldots\right) = 
\frac{1}{2} R\, \tilde U_{\text{max.}}^2 - \frac{1}{8R}\log \tilde U_{\text{max.}} 
+\text{finite}.
\ee
The Legendre transformation may be implemented in analogy with the
fundamental string case. One defines variables $Y^I$ such that
\bsp \label{YcoordsM}
&\frac{d\tilde U^2}{\tilde U^2}+\frac{1}{4}d\O_4^2 = 
\frac{1}{4}\left(\frac{dV^2}{V^2}+d\O_4^2\right) =
\frac{dY^I d{Y}^I}{4Y^2},\\
&Y^I = V \hat \T^I, \quad \hat \T^I
\hat \T^I=1, \quad I=1,\ldots,5,  
\end{split}
\ee
so that $V=\tilde U^2$. Then the action is regularized via
\bsp\label{legendreM}
S_{M2\,\text{reg.}} &= S_{M2} - \int d^2\t\, d\s \,
\p_\s \left(Y^I \frac{\d {\cal L}}{\d \p_\s Y^I} \right)\\
& = S_{M2} - \int d^2\t\, Y^I \frac{\d {\cal L}}{\d \p_\s Y^I} \Biggr|_{\p\S},
\end{split}
\ee
where the boundary surface $\p \S$ is parameterized by the two $\t$
coordinates. This removes the leading divergence\footnote{As in the
  string case the boundary term in the Legendre transformation
  also contributes a finite term.} from
(\ref{M2act}). We then find that the expectation value of our Wilson
surface is
\be\label{M}
\la W_{\text{surface}}\ra = \tilde{\cal V}\,
\exp\left(-S_{M2\,\text{reg.}}\right) =\tilde{\cal V}\,
\exp \left( \frac{\pi NR_6}{R} \log \tilde U_{\text{max.}}\right) \cdot
(\text{finite}),
\ee
where we have included an unknown semi-classical prefactor $\tilde
{\cal V}$ as in the string case. Using the fact that $\tilde U \sim
U^{1/2}$ (see footnote \ref{foot1}), and the identification
$g^2/(8\pi^2)=R_6$, the string (\ref{string}) and M-theory (\ref{M})
results match, at least for the exponential factor containing the
logarithmic divergence and for the finite part, since the action
(\ref{membact}) is equivalent to the string action (\ref{strgact}) with $p=4$. The
prefactor is sensitive to fluctuations of the classical solutions and
could be different between M and string theory.

A similar logarithmic divergence is seen in the spherical
M2-brane solution in $AdS_7\times S^4$ presented in \cite{Berenstein:1998ij} (section
5). The metric on $AdS_7$ is taken as $dU^2/U^2 + U^2 dx_i^2$ and the
solution is
\be
x^i(U,\theta,\phi) = \sqrt{R^2-U^{-2}} \left(\sin\theta\cos\phi,\,
\sin\theta\sin\phi,\,\cos\theta,0,0,0\right).
\ee
The M2-brane action then evaluates to
\be\label{M2lift}
S_{M2} = 4N\left(R^2 U_{\text{max.}}^2 
- \log \left(2\,R\,U_{\text{max.}}\right)-\frac{1}{2}\right).
\ee
Again, the regularization procedure (\ref{legendreM}) removes the
leading term. Then we are left with a logarithmic divergence, causing
the expectation value to be scale-dependent. Since M-theory on
$AdS_7\times S^4$ is dual to the $(2,0)$ $d=6$ CFT, this scale
dependence might appear surprising.  As mentioned in the introduction,
it has been understood as a conformal anomaly suffered by sub-manifold
observables corresponding to $k$-branes for even $k$ (such as Wilson
surfaces) in CFT's
\cite{Graham:1999pm,Henningson:1999xi,Gustavsson:2003hn,Gustavsson:2004gj}.
We may therefore interpret the logarithmic divergence in
(\ref{string}) as arising in $d=5$ SYM, via dimensional reduction,
from this $d=6$ anomaly. It is remarkable that, given the recent
speculations that the $(2,0)$ CFT in six dimensions might actually
also be described by five-dimensional SYM
\cite{Lambert:2010iw,Bolognesi:2011nh,Bolognesi:2011rq,Douglas:2010iu,Gustavsson:2011ur},
we are able to recover this exponential term by summing planar ladder
diagrams directly in $d=5$ SYM, see section \ref{gauge}.

\subsubsection{Generalization to smooth, closed contours}
\label{stringgen}

Before turning to the gauge theory analysis, we remark that the
coefficient of the $\log$ divergence is known for general shaped
Wilson surfaces. It is given by the rigid string action
\cite{Berenstein:1998ij,Graham:1999pm}
\be
S_{\text{rigid}} = \frac{N}{4\pi}\int d^2\t\, \sqrt{h} \left( \nabla^2 X^i \right)^2,
\ee
where the embedding of the surface in $\bR^6$
is given by $X^i(\t_1,\t_2)$ and where $h_{\a\b} = \p_\a X^i \p_\b X^i$ is
the induced metric. If we take $\t_1$ to parametrize the 5-d Wilson loop
contour $x^I(\t_1)$ by arclength, and $\t_2$ to be the trvially embedded
periodically identified M-theory direction, 
\be
X^i(\t_1,\t_2) = \left(x^I(\t_1), \t_2\right), \qquad I=1,\ldots,5,
\ee
this reduces to
\be\label{blahp}
S_{\text{rigid}} = \frac{N R_6}{2} \int d\t_1 \left(\p^2_{\t_1} x^I\right)^2.
\ee 

We may now consider a string worldsheet in the $D4$-brane geometry
with embedding functions $x^I(\t_1,U)$ such that $x^I(\t_1,\infty) =
x^I(\t_1)$, i.e. a string describing the Wilson loop. The string action reads
\be
S = \frac{1}{2\pi}\int d\t_1 \,dU \sqrt{\dot x^2\left(1 +
  x'^2\frac{U^3}{C_4^2}\right)
-\left(\dot x \cdot x'\right)^2 \frac{U^3}{C_4^2}} ,
\ee 
where we take $\dot x^I \equiv \p_{\t_1}x^I$ and ${x'}^I \equiv \p_U x^I$.
One may verify that for large $U$, the following ansatz satisfies the
equations of motion
\be
x^I(\t_1,U) \simeq x^I(\t_1) + \frac{C_4^2}{U} \ddot x^I(\t_1),
\ee
where we remind the reader that $x^I(\t_1)$ is parametrized by
arclength. Evaluating the action for large $U$ one obtains
\be
S =\frac{1}{2\pi} L\,U_{\text{max.}} - \frac{g^2 N}{32\pi^2} 
\left(\int d\t_1 \, \ddot x^2\right)
\log U_{\text{max.}}+ \text{subleading}  ,
\ee
where $L$ is the arclength of the curve and where we have used the
fact that $C_4^2 = g^2N/(8\pi)$. The Legendre transformation removes
the leading divergence. Then, using the dictionary between
$M$-theoretic and string theoretic quantities as in section
\ref{m2surf}, we find that we have an exact match with (\ref{blahp}).

\section{Gauge theory analysis}
\label{gauge}

\subsection{Exponential factor from planar diagrams}
\label{gauge2}

In this section we will recover (\ref{string}) by summing ladder
diagrams in Euclidean five-dimensional SYM, finding the exact
exponent. The Wilson loop is defined in the gauge theory as follows
\be\label{wl}
W = \frac{1}{N} \Tr {\cal P} \exp \oint d\t \left( i\dot x^\m A_\m +
|\dot x| \Theta^I \Phi_I\right),
\ee
where $\Phi_I$ with $I=1,\ldots,5$ are the five real scalars of the
SYM theory, $A_\m$ is the gauge field, and ${\cal P}$
indicates path ordering. The trace will be taken in the fundamental
representation of the gauge group $SU(N)$, whose generators $T^a$ are
normalized by $\Tr(T^aT^b)=\d^{ab}/2$.

We take $x^\m=R\,(\cos\t,\,\sin\t,\,0,0,0)$ and $\Theta^I =
\text{const.}$. Using dimensional reduction from ${\cal N}=1$, $d=10$
SYM to $d=2\o$ dimensions \cite{Erickson:2000af}, the Feynman-gauge
propagators are as follows
\be
\la A_\m^a (x) \,A_\n^b(0)\ra = \G(\o-1)\frac{g^2}{4\pi^\o}
\frac{\d^{ab}\d_{\m\n}}{x^{2\o-2}},\quad
\la \Phi_I^a (x) \,\Phi_J^b(0)\ra = \G(\o-1)\frac{g^2}{4\pi^\o}
\frac{\d^{ab}\d_{IJ}}{x^{2\o-2}}.
\ee
The fundamental object of interest is the loop-to-loop propagator
which refers to the 1-gluon $+$ 1-scalar exchange between two
locations $x_1^\m=x^\m(\t_1)$ and $x_2^\m=x^\m(\t_2)$ on the Wilson
loop contour
\be\label{loop2loop}
\Bigl\la \Bigl( i\dot x^\m A^a_\m +
|\dot x| \Theta^I \Phi_I^a\Bigr) (x_1) 
\Bigl( i\dot x^\m A_\m^b +
|\dot x| \Theta^I \Phi_I^b\Bigr) (x_2)\Bigr\ra =
\frac{g^2\G(\o-1)}{\pi^\o R^{2\o-4}} 
\frac{
\d_{ab}\,2^{1-2\o}}{\sin^{2\o-4}\frac{\t_{1}-\t_2}{2}}.  
\ee
We see that an integral over $\t_1$ and $\t_2$ will produce
a $1/(5-2\o) \equiv 1/\e$ divergence as $\t_2$ approaches
$\t_1$. Alternatively, we could set $2\o=5$ and regulate this
divergence using point-splitting, i.e. by cutting the integral off at
$\t_1-\t_2 = \d$. We will continue by considering both regularizations.   

\begin{figure}\label{fig:outermost}
\begin{center}
\includegraphics*[width=2in]{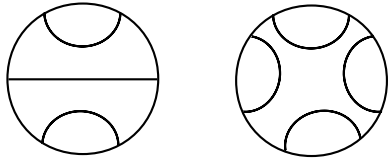}
\end{center}
\caption{Examples of $(q,k)$-graphs. On the left a $(2,3)$-graph is
  shown; on the right a $(4,4)$-graph is shown.}
\end{figure}

Summing the planar rainbow/ladders was done for the $d=4$ theory in
\cite{Erickson:2000af}, the difference here is that our loop-to-loop
propagator is not constant. For this reason we will need to take a
closer look at the diagrams. It was shown in \cite{Erickson:2000af},
that the number $N_k$ of planar rainbow/ladder diagrams with $k$
propagators is given by the $k^{\text{th}}$ Catalan number
\be
N_k = C_k = \frac{(2k)!}{k!(k+1)!}.
\ee
It will be important for our considerations to further subdivide the
diagrams by the number of outermost propagators they contain, see
figure \ref{fig:outermost}. 
An outermost propagator is defined as a
propagator which encloses, between itself and the Wilson loop contour,
no other propagators. Let us denote the diagrams with $q$ outermost
propagators and $k$ total propagators as $(q,k)$-graphs. It is obvious
that $q\in[2,k]$. The multiplicity $M_{q,k}$ of the $(q,k)$-graphs
is\footnote{There is only one $(1,1)$-graph -- the only graph with one
  propagator -- and so the formula applies only to diagrams with two
  or more propagators.} 
\be\label{Mqk}
M_{q,k} = \frac{2\,k!(k-2)!}{q!(q-2)!(k-q)!(k-q+1)!},
\ee
and one can verify that 
\be
\sum_{q=2}^k M_{q,k} = N_k,
\ee
as it must.

One finds that there is an association, at a given loop-level
(i.e. fixed $k$), between the divergence of a graph and its $q$-value,
the maximum divergence coming from the maximum value of $q$,
i.e. $q=k$. Indeed we find that 
\be
(q,k)\text{-graph} \propto \begin{cases}
(\frac{1}{\e})^q + \text{subleading},\qquad&\text{dim. red.}\\
(\log\d)^q + \text{subleading},\qquad&\text{point-split.}
\end{cases}.
\ee
We would therefore like to sum-up the most divergent diagrams, the
$(k,k)$-graphs. It turns-out that the integration associated to these
graphs has a simple closed form at the leading order in
small-$\e$. The integral is as follows
\be
I_{k,k} = 
\frac{2\pi}{2k} \int_0^{2\pi} d\theta_{2k-1}
\int_0^{\theta_{2k-1}}d\theta_{2k-2}
\ldots \int_0^{\theta_2} d\theta_1\,
\frac{1}{\sin^{1-\e}\frac{\theta_1}{2}} \prod_{j=2}^{k}
\frac{1}{\sin^{1-\e}\frac{\theta_{2j-1}-\theta_{2j-2}}{2}},
\ee 
in dimensional reduction, or
\be
I_{k,k} = 
\frac{2\pi}{2k} \int_\d^{2\pi-\d} d\theta_{2k-1}
\int_\d^{\theta_{2k-1}-\d}d\theta_{2k-2}
\ldots \int_\d^{\theta_2-\d} d\theta_1\,
\frac{1}{\sin\frac{\theta_1}{2}} \prod_{j=2}^{k}
\frac{1}{\sin\frac{\theta_{2j-1}-\theta_{2j-2}}{2}},
\ee 
in point-splitting regularization,
so that the contribution in perturbation theory is
\be
\left(\frac{g^2N\,\G(\o-1)}{2^{2\o}\pi^\o R^{2\o-4}}\right)^k I_{k,k}\, M_{k,k} ,
\ee 
where we have included the planar colour factor $(N/2)^k$. We find
that
\bsp
I_{k,k} = &\frac{2\pi}{2k} \left(2\begin{cases}\frac{1}{\e}\\
-\log\d \end{cases}\right)^k \,
\frac{(2\pi)^{k-1}}{(k-1)!}+\text{subleading},
\end{split}
\ee
and therefore, using the fact that $M_{k,k}=2$, we have that the
contribution is\footnote{In the $k=1$ case the integral $I_{1,1}$ has
  a compensatory factor of 2, accounting for the fact that $M_{1,1}=1$.}
\be
\frac{1}{k!}\left(\frac{g^2N}{16\pi R}
\right)^k \begin{cases}\left(\frac{1}{\e}\right)^k\\
\left(-\log\d\right)^k \end{cases},
\ee
which therefore sums to
\be\label{expsum}
 \exp\Biggl(\frac{g^2N}{16\pi R} \begin{cases}\frac{1}{\e}\\
-\log\d \end{cases}
\Biggr),
\ee
and identifying the UV cut-off $U_{\text{max.}}$ with $\exp(1/\e)$ (or $1/\d$,
in the point-split case), we find
that we have recovered exactly the exponential factor found from the
string theory analysis (\ref{string}).

We have neglected the subleading divergences, which amount to lower
powers of $\e^{-1}$'s or $\log\d$'s at each loop-order. Since these must sum-up to
something less divergent than the exponential factor (\ref{expsum}),
it is natural to associate them with the prefactor ${\cal V}$
appearing in (\ref{string}). It would be interesting to find a way to
verify this idea from the string or M-theoretic perspective. These
subleading divergences are of course scheme-dependent, as a
redefinition of $\e$ can be used to tune the subleading
coefficients.

\subsection{Interacting diagrams at one loop}
\label{int}

\begin{figure}
\begin{center}
\includegraphics*[height=1.0in]{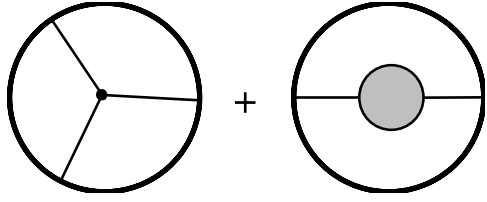}
\end{center}
\caption{The one-loop, non-ladder/rainbow diagrams at ${\cal O}(g^4)$.
  One the left is $\S_3$, and on the right $\S_2$.
  Internal solid lines refer to scalar and gauge fields, while the
  greyed-in bubble represents the one-loop correction to the
  propagator.}
\label{fig:iv}
\end{figure}

The one-loop analysis of Wilson loops in SYM theories obtainable via
dimensional reduction from ${\cal N}=1$, $d=10$ SYM appeared
originally in \cite{Erickson:2000af}. In \cite{Young:2008ed}, a more
general presentation was made of the same results. The two diagrams to
be considered are shown in figure \ref{fig:iv}.
The trivalent graph $\S_3$ is built from the following function, a result of
integrating over the position of the triple-vertex
\bsp G(x_1,x_2,x_3) = \frac{\G(2\o-3)}{2^6 \pi^{2\o}} \int_0^1
d\alpha\,d\beta\,d\gamma\,(\alpha\beta\gamma)^{\omega-2}
\delta(1-\alpha-\beta-\gamma)\,\\
\times \frac{1} {\bigl[ \a\b(x_1-x_2)^2 + \b\g(x_2-x_3)^2 +
\a\g(x_1-x_3)^2 \bigr]^{2\o-3}},
\end{split}
\ee
while the one-loop-corrected propagator graph $\S_2$ may be added to
$\S_3$ using an integration-by-parts trick \cite{Erickson:2000af}
\be\label{2l} \S_3 + \S_2 = -\frac{g^4N^2}{4} \oint d\t_1 \,d\t_2
\,d\t_3 \, \e(\t_1\,\t_2\,\t_3)\, \Biggl[ D(\t_1,\t_3)\,\dot x_2
\cdot \p_{x_1} G - \p_{\t_1} \biggl(D(\t_1,\t_3)\,G \biggr) \Biggr],
\ee
where $\e(\t_1\,\t_2\,\t_3)$ refers to antisymmetric path-ordering
given by +1 for $\t_1 > \t_2 > \t_3$ and totally antisymmetric in the
$\t_i$, and where $D(\t_1,\t_2) = |\dot x_1||\dot x_2| - \dot x_1
\cdot \dot x_2$ is the numerator of the loop-to-loop propagator
(\ref{loop2loop}). Plugging in the circular contour, we find
\bsp\label{finite}
&\S_2+\S_3 = 
(4-2\o)\frac{g^4N^2\,\G(2\o-3)}{3\cdot2^{2\o+3}\pi^{2\o}R^{4\o-8}}  \int_0^1
d\alpha\,d\beta\,d\gamma\,
\delta(1-\alpha-\beta-\gamma)\\
&\times
\oint d\t_1 \,d\t_2
\,d\t_3 \, \e(\t_1\,\t_2\,\t_3)\,
\frac{\left(\alpha\beta\gamma\right)^{\o-2}\left(\sin\t_{13}+\sin\t_{32}+\sin\t_{21}\right)}
{\bigl[\a\b(1-\cos\t_{12})+\b\g(1-\cos\t_{23})+\a\g(1-\cos\t_{13})\bigr]^{2\o-3}}.
\end{split}
\ee
For $d=2\o=5$ this is a finite\footnote{\label{ftft}The convergence of
  this integral was discussed in \cite{Young:2008ed}, section 2.3. One
  takes one of the three Feynman parameters (e.g. $\g$) near
  zero. Integration-by-parts on one of the $\t_i$ shows that
  divergences in the $\t_i$ integration cancel. The Feynman parameter
  integration is then seen to be convergent only for $2\o < 6$.}
integral which can be evaluated numerically. Stripping-off the factor
to the left of the integral signs in (\ref{finite}), we find a value
of $-499 \pm 3$ using a standard Monte-Carlo integration of $10^9$
steps. We find however that for $d\geq 6$, the integral is divergent.

\subsection{Generalizing to any smooth closed curve}
\label{shape}

As explained in section \ref{stringgen}, the coefficient of the
logarithmic divergence is given from string theory for a general
closed contour by the following expression
\be\label{genM}
S_{\text{rigid}} = \frac{N R_6}{2} \int d\t_1 \left(\p^2_{\t_1} x^I\right)^2.
\ee
In fact we can generalize the gauge theory analysis of the previous
section by noting that the leading divergence comes from the
propagators in the $(k,k)$-graphs pinching their ends
together. Analyzing the loop-to-loop propagator in this pinched limit,
one finds (again, using arclength parametrization)
\be
\lim_{\t_1\to\t_2} P(\t_1,\t_2)=
\lim_{\t_1\to\t_2}\frac{|\dot x_1||\dot x_2| - \dot x_1\cdot \dot x_2}{(x_1-x_2)^{2\o-2}} 
\simeq
\frac{\ddot x_1^2}{2\,(\t_1-\t_2)^{1-\e}} .
\ee 
The associated path-ordered integral (we denote the circumference of
the closed curve by $L$)\footnote{The leading factor of 2 in
  (\ref{thingy}) comes from $M_{k,k}$, see (\ref{Mqk}).}
\be\label{thingy}
2\left(\frac{g^2N}{16\pi^2}\right)^k\int_0^L d\t_{2k}\int_0^{\t_{2k}} d\t_{2k-1} \cdots \int_0^{\t_2}d\t_1
\, P(\t_1,\t_2) P(\t_3,\t_4) \cdots P(\t_{2k-1},\t_{2k})
\ee
has a leading divergence in the pinched limit given by 
\bsp
&2\left(\frac{g^2N}{16\pi^2}\right)^k\left(\frac{1}{2\e}\right)^k \int_0^L d\t_k
\int_0^{\t_{k}}d\t_{k-1}\cdots \int_0^{\t_2}d\t_1\,
\ddot x_k^2 \cdots \ddot x_1^2\\
& \qquad\qquad\qquad\qquad\qquad\qquad
= \left(\frac{g^2N}{32\pi^2\e}\right)^k\frac{2}{k!}
\left(\int_0^L d\t \,\ddot x^2(\t) \right)^k.
\end{split}
\ee
We therefore find that the leading divergence exponentiates
\be
\log \la W \ra = \frac{g^2N}{32\pi^2\e} \int_0^L d\t \, \ddot x^2 + \text{subleading},
\ee
giving, upon using $R_6 =g^2/(8\pi^2)$ and accounting for the fact
that the $\log$ of the M-theory coordinate $\tilde U$ shoud be
identified with $1/(2\e)$ (as established for the circle in sections
\ref{m2surf} and \ref{gauge2}), exactly (\ref{genM}). We note that a very
similar exponentiation was found in the four-dimensional context in
\cite{Pestun:2002mr}, where correlators of Wilson loops with operators
of large R-charge were compared to string theory. 

The interacting diagrams at ${\cal O}(g^4)$ remain finite for general
smooth contours. The arguments given in footnote \ref{ftft} continue
to apply to well-behaved $D(\t_1,\t_2)$ (see (\ref{2l})), i.e. for
closed contours free of cusps.

\section*{Acknowledgments}

We thank Abhishek Agarwal for discussions and comments. This work was
supported by FNU through grant number 272-08-0329.

\bibliography{5dloops}%

\providecommand{\href}[2]{#2}\begingroup\raggedright\begin{thebibliography}{10}

\bibitem{Maldacena:1998im}
J.~M. Maldacena, ``{Wilson loops in large N field theories},''
  \href{http://dx.doi.org/10.1103/PhysRevLett.80.4859}{{\em Phys.Rev.Lett.}
  {\bf 80} (1998)  4859--4862}, \href{http://arxiv.org/abs/hep-th/9803002}{{\tt
  arXiv:hep-th/9803002 [hep-th]}}.

\bibitem{Rey:1998ik}
S.-J. Rey and J.-T. Yee, ``{Macroscopic strings as heavy quarks in large N
  gauge theory and anti-de Sitter supergravity},''
  \href{http://dx.doi.org/10.1007/s100520100799}{{\em Eur.Phys.J.} {\bf C22}
  (2001)  379--394}, \href{http://arxiv.org/abs/hep-th/9803001}{{\tt
  arXiv:hep-th/9803001 [hep-th]}}.

\bibitem{Erickson:2000af}
J.~Erickson, G.~Semenoff, and K.~Zarembo, ``{Wilson loops in N=4 supersymmetric
  Yang-Mills theory},''
  \href{http://dx.doi.org/10.1016/S0550-3213(00)00300-X}{{\em Nucl.Phys.} {\bf
  B582} (2000)  155--175}, \href{http://arxiv.org/abs/hep-th/0003055}{{\tt
  arXiv:hep-th/0003055 [hep-th]}}.

\bibitem{Drukker:2000rr}
N.~Drukker and D.~J. Gross, ``{An Exact prediction of N=4 SUSYM theory for
  string theory},'' \href{http://dx.doi.org/10.1063/1.1372177}{{\em
  J.Math.Phys.} {\bf 42} (2001)  2896--2914},
\href{http://arxiv.org/abs/hep-th/0010274}{{\tt arXiv:hep-th/0010274
  [hep-th]}}.

\bibitem{Drukker:2008zx}
N.~Drukker, J.~Plefka, and D.~Young, ``{Wilson loops in 3-dimensional N=6
  supersymmetric Chern-Simons Theory and their string theory duals},''
  \href{http://dx.doi.org/10.1088/1126-6708/2008/11/019}{{\em JHEP} {\bf 0811}
  (2008)  019},
\href{http://arxiv.org/abs/0809.2787}{{\tt arXiv:0809.2787 [hep-th]}}.

\bibitem{Drukker:2009hy}
N.~Drukker and D.~Trancanelli, ``{A Supermatrix model for N=6 super
  Chern-Simons-matter theory},''
  \href{http://dx.doi.org/10.1007/JHEP02(2010)058}{{\em JHEP} {\bf 1002} (2010)
   058},
\href{http://arxiv.org/abs/0912.3006}{{\tt arXiv:0912.3006 [hep-th]}}.

\bibitem{Chen:2008bp}
B.~Chen and J.-B. Wu, ``{Supersymmetric Wilson Loops in N=6 Super
  Chern-Simons-matter theory},''
  \href{http://dx.doi.org/10.1016/j.nuclphysb.2009.09.015}{{\em Nucl.Phys.}
  {\bf B825} (2010)  38--51},
\href{http://arxiv.org/abs/0809.2863}{{\tt arXiv:0809.2863 [hep-th]}}.

\bibitem{Rey:2008bh}
S.-J. Rey, T.~Suyama, and S.~Yamaguchi, ``{Wilson Loops in Superconformal
  Chern-Simons Theory and Fundamental Strings in Anti-de Sitter Supergravity
  Dual},'' \href{http://dx.doi.org/10.1088/1126-6708/2009/03/127}{{\em JHEP}
  {\bf 0903} (2009)  127},
\href{http://arxiv.org/abs/0809.3786}{{\tt arXiv:0809.3786 [hep-th]}}.

\bibitem{Pestun:2007rz}
V.~Pestun, ``{Localization of gauge theory on a four-sphere and supersymmetric
  Wilson loops},''
\href{http://arxiv.org/abs/0712.2824}{{\tt arXiv:0712.2824 [hep-th]}}.

\bibitem{Kapustin:2009kz}
A.~Kapustin, B.~Willett, and I.~Yaakov, ``{Exact Results for Wilson Loops in
  Superconformal Chern-Simons Theories with Matter},''
  \href{http://dx.doi.org/10.1007/JHEP03(2010)089}{{\em JHEP} {\bf 1003} (2010)
   089},
\href{http://arxiv.org/abs/0909.4559}{{\tt arXiv:0909.4559 [hep-th]}}.

\bibitem{Marino:2009jd}
M.~Marino and P.~Putrov, ``{Exact Results in ABJM Theory from Topological
  Strings},'' \href{http://dx.doi.org/10.1007/JHEP06(2010)011}{{\em JHEP} {\bf
  1006} (2010)  011},
\href{http://arxiv.org/abs/0912.3074}{{\tt arXiv:0912.3074 [hep-th]}}.

\bibitem{Drukker:2010nc}
N.~Drukker, M.~Marino, and P.~Putrov, ``{From weak to strong coupling in ABJM
  theory},'' \href{http://dx.doi.org/10.1007/s00220-011-1253-6}{{\em
  Commun.Math.Phys.} {\bf 306} (2011)  511--563},
\href{http://arxiv.org/abs/1007.3837}{{\tt arXiv:1007.3837 [hep-th]}}.

\bibitem{Drukker:2005kx}
N.~Drukker and B.~Fiol, ``{All-genus calculation of Wilson loops using
  D-branes},'' \href{http://dx.doi.org/10.1088/1126-6708/2005/02/010}{{\em
  JHEP} {\bf 0502} (2005)  010},
  \href{http://arxiv.org/abs/hep-th/0501109}{{\tt arXiv:hep-th/0501109
  [hep-th]}}.

\bibitem{Gomis:2006sb}
J.~Gomis and F.~Passerini, ``{Holographic Wilson Loops},''
  \href{http://dx.doi.org/10.1088/1126-6708/2006/08/074}{{\em JHEP} {\bf 0608}
  (2006)  074}, \href{http://arxiv.org/abs/hep-th/0604007}{{\tt
  arXiv:hep-th/0604007 [hep-th]}}.

\bibitem{Gomis:2006im}
J.~Gomis and F.~Passerini, ``{Wilson Loops as D3-Branes},''
  \href{http://dx.doi.org/10.1088/1126-6708/2007/01/097}{{\em JHEP} {\bf 0701}
  (2007)  097}, \href{http://arxiv.org/abs/hep-th/0612022}{{\tt
  arXiv:hep-th/0612022 [hep-th]}}.

\bibitem{Hartnoll:2006is}
S.~A. Hartnoll and S.~Kumar, ``{Higher rank Wilson loops from a matrix
  model},'' \href{http://dx.doi.org/10.1088/1126-6708/2006/08/026}{{\em JHEP}
  {\bf 0608} (2006)  026}, \href{http://arxiv.org/abs/hep-th/0605027}{{\tt
  arXiv:hep-th/0605027 [hep-th]}}.

\bibitem{Hartnoll:2006ib}
S.~A. Hartnoll, ``{Two universal results for Wilson loops at strong
  coupling},'' \href{http://dx.doi.org/10.1103/PhysRevD.74.066006}{{\em
  Phys.Rev.} {\bf D74} (2006)  066006},
  \href{http://arxiv.org/abs/hep-th/0606178}{{\tt arXiv:hep-th/0606178
  [hep-th]}}.

\bibitem{Yamaguchi:2006tq}
S.~Yamaguchi, ``{Wilson loops of anti-symmetric representation and
  D5-branes},'' \href{http://dx.doi.org/10.1088/1126-6708/2006/05/037}{{\em
  JHEP} {\bf 0605} (2006)  037},
  \href{http://arxiv.org/abs/hep-th/0603208}{{\tt arXiv:hep-th/0603208
  [hep-th]}}.

\bibitem{D'Hoker:2007fq}
E.~D'Hoker, J.~Estes, and M.~Gutperle, ``{Gravity duals of half-BPS Wilson
  loops},'' \href{http://dx.doi.org/10.1088/1126-6708/2007/06/063}{{\em JHEP}
  {\bf 0706} (2007)  063},
\href{http://arxiv.org/abs/0705.1004}{{\tt arXiv:0705.1004 [hep-th]}}.

\bibitem{Yamaguchi:2006te}
S.~Yamaguchi, ``{Bubbling geometries for half BPS Wilson lines},''
  \href{http://dx.doi.org/10.1142/S0217751X07035070}{{\em Int.J.Mod.Phys.} {\bf
  A22} (2007)  1353--1374},
\href{http://arxiv.org/abs/hep-th/0601089}{{\tt arXiv:hep-th/0601089
  [hep-th]}}.

\bibitem{Lunin:2006xr}
O.~Lunin, ``{On gravitational description of Wilson lines},''
  \href{http://dx.doi.org/10.1088/1126-6708/2006/06/026}{{\em JHEP} {\bf 0606}
  (2006)  026},
\href{http://arxiv.org/abs/hep-th/0604133}{{\tt arXiv:hep-th/0604133
  [hep-th]}}.

\bibitem{Itzhaki:1998dd}
N.~Itzhaki, J.~M. Maldacena, J.~Sonnenschein, and S.~Yankielowicz,
  ``{Supergravity and the large N limit of theories with sixteen
  supercharges},'' \href{http://dx.doi.org/10.1103/PhysRevD.58.046004}{{\em
  Phys.Rev.} {\bf D58} (1998)  046004},
  \href{http://arxiv.org/abs/hep-th/9802042}{{\tt arXiv:hep-th/9802042
  [hep-th]}}.

\bibitem{Agarwal:2009up}
A.~Agarwal and D.~Young, ``{Supersymmetric Wilson Loops in Diverse
  Dimensions},'' \href{http://dx.doi.org/10.1088/1126-6708/2009/06/063}{{\em
  JHEP} {\bf 0906} (2009)  063}, \href{http://arxiv.org/abs/0904.0455}{{\tt
  arXiv:0904.0455 [hep-th]}}.

\bibitem{Zarembo:2002an}
K.~Zarembo, ``{Supersymmetric Wilson loops},''
  \href{http://dx.doi.org/10.1016/S0550-3213(02)00693-4}{{\em Nucl.Phys.} {\bf
  B643} (2002)  157--171}, \href{http://arxiv.org/abs/hep-th/0205160}{{\tt
  arXiv:hep-th/0205160 [hep-th]}}.

\bibitem{Drukker:1999zq}
N.~Drukker, D.~J. Gross, and H.~Ooguri, ``{Wilson loops and minimal
  surfaces},'' \href{http://dx.doi.org/10.1103/PhysRevD.60.125006}{{\em
  Phys.Rev.} {\bf D60} (1999)  125006},
  \href{http://arxiv.org/abs/hep-th/9904191}{{\tt arXiv:hep-th/9904191
  [hep-th]}}.

\bibitem{Lambert:2010iw}
N.~Lambert, C.~Papageorgakis, and M.~Schmidt-Sommerfeld, ``{M5-Branes,
  D4-Branes and Quantum 5D super-Yang-Mills},''
  \href{http://dx.doi.org/10.1007/JHEP01(2011)083}{{\em JHEP} {\bf 1101} (2011)
   083}, \href{http://arxiv.org/abs/1012.2882}{{\tt arXiv:1012.2882 [hep-th]}}.

\bibitem{Douglas:2010iu}
M.~R. Douglas, ``{On D=5 super Yang-Mills theory and (2,0) theory},''
  \href{http://dx.doi.org/10.1007/JHEP02(2011)011}{{\em JHEP} {\bf 1102} (2011)
   011},
\href{http://arxiv.org/abs/1012.2880}{{\tt arXiv:1012.2880 [hep-th]}}.

\bibitem{Graham:1999pm}
C.~Graham and E.~Witten, ``{Conformal anomaly of submanifold observables in AdS
  / CFT correspondence},''
  \href{http://dx.doi.org/10.1016/S0550-3213(99)00055-3}{{\em Nucl.Phys.} {\bf
  B546} (1999)  52--64},
\href{http://arxiv.org/abs/hep-th/9901021}{{\tt arXiv:hep-th/9901021
  [hep-th]}}.

\bibitem{Henningson:1999xi}
M.~Henningson and K.~Skenderis, ``{Weyl anomaly for Wilson surfaces},'' {\em
  JHEP} {\bf 9906} (1999)  012,
\href{http://arxiv.org/abs/hep-th/9905163}{{\tt arXiv:hep-th/9905163
  [hep-th]}}.

\bibitem{Gustavsson:2003hn}
A.~Gustavsson, ``{On the Weyl anomaly of Wilson surfaces},'' {\em JHEP} {\bf
  0312} (2003)  059,
\href{http://arxiv.org/abs/hep-th/0310037}{{\tt arXiv:hep-th/0310037
  [hep-th]}}.

\bibitem{Gustavsson:2004gj}
A.~Gustavsson, ``{Conformal anomaly of Wilson surface observables: A Field
  theoretical computation},''
  \href{http://dx.doi.org/10.1088/1126-6708/2004/07/074}{{\em JHEP} {\bf 0407}
  (2004)  074},
\href{http://arxiv.org/abs/hep-th/0404150}{{\tt arXiv:hep-th/0404150
  [hep-th]}}.

\bibitem{Bolognesi:2011nh}
S.~Bolognesi and K.~Lee, ``{Instanton Partons in 5-dim SU(N) Gauge Theory},''
  \href{http://dx.doi.org/10.1103/PhysRevD.84.106001}{{\em Phys.Rev.} {\bf D84}
  (2011)  106001},
\href{http://arxiv.org/abs/1106.3664}{{\tt arXiv:1106.3664 [hep-th]}}.

\bibitem{Bolognesi:2011rq}
S.~Bolognesi and K.~Lee, ``{1/4 BPS String Junctions and $N^3$ Problem in 6-dim
  (2,0) Superconformal Theories},''
\href{http://arxiv.org/abs/1105.5073}{{\tt arXiv:1105.5073 [hep-th]}}.

\bibitem{Gustavsson:2011ur}
A.~Gustavsson, ``{A preliminary test of Abelian D4-M5 duality},''
\href{http://arxiv.org/abs/1111.6339}{{\tt arXiv:1111.6339 [hep-th]}}.

\bibitem{Jevicki:1998ub}
A.~Jevicki, Y.~Kazama, and T.~Yoneya, ``{Generalized conformal symmetry in
  D-brane matrix models},''
  \href{http://dx.doi.org/10.1103/PhysRevD.59.066001}{{\em Phys.Rev.} {\bf D59}
  (1999)  066001},
\href{http://arxiv.org/abs/hep-th/9810146}{{\tt arXiv:hep-th/9810146
  [hep-th]}}.

\bibitem{Frenkel:1984pz}
J.~Frenkel and J.~Taylor, ``{NONABELIAN EIKONAL EXPONENTIATION},''
\href{http://dx.doi.org/10.1016/0550-3213(84)90294-3}{{\em Nucl.Phys.} {\bf
  B246} (1984)  231}.

\bibitem{Gatheral:1983cz}
J.~Gatheral, ``{EXPONENTIATION OF EIKONAL CROSS-SECTIONS IN NONABELIAN GAUGE
  THEORIES},''
\href{http://dx.doi.org/10.1016/0370-2693(83)90112-0}{{\em Phys.Lett.} {\bf
  B133} (1983)  90}.

\bibitem{Brandt:1981kf}
R.~A. Brandt, F.~Neri, and M.-a. Sato, ``{Renormalization of Loop Functions for
  All Loops},''
\href{http://dx.doi.org/10.1103/PhysRevD.24.879}{{\em Phys.Rev.} {\bf D24}
  (1981)  879}.

\bibitem{Polyakov:1980ca}
A.~M. Polyakov, ``{Gauge Fields as Rings of Glue},''
\href{http://dx.doi.org/10.1016/0550-3213(80)90507-6}{{\em Nucl.Phys.} {\bf
  B164} (1980)  171--188}.

\bibitem{Corrado:1999pi}
R.~Corrado, B.~Florea, and R.~McNees, ``{Correlation functions of operators and
  Wilson surfaces in the d = 6, (0,2) theory in the large N limit},''
  \href{http://dx.doi.org/10.1103/PhysRevD.60.085011}{{\em Phys.Rev.} {\bf D60}
  (1999)  085011},
\href{http://arxiv.org/abs/hep-th/9902153}{{\tt arXiv:hep-th/9902153
  [hep-th]}}.

\bibitem{Kanitscheider:2008kd}
I.~Kanitscheider, K.~Skenderis, and M.~Taylor, ``{Precision holography for
  non-conformal branes},''
  \href{http://dx.doi.org/10.1088/1126-6708/2008/09/094}{{\em JHEP} {\bf 0809}
  (2008)  094},
\href{http://arxiv.org/abs/0807.3324}{{\tt arXiv:0807.3324 [hep-th]}}.

\bibitem{Wiseman:2008qa}
T.~Wiseman and B.~Withers, ``{Holographic renormalization for coincident
  Dp-branes},'' \href{http://dx.doi.org/10.1088/1126-6708/2008/10/037}{{\em
  JHEP} {\bf 0810} (2008)  037},
\href{http://arxiv.org/abs/0807.0755}{{\tt arXiv:0807.0755 [hep-th]}}.

\bibitem{Chu:2008xg}
C.-S. Chu and D.~Giataganas, ``{UV-divergences of Wilson Loops for
  Gauge/Gravity Duality},''
  \href{http://dx.doi.org/10.1088/1126-6708/2008/12/103}{{\em JHEP} {\bf 0812}
  (2008)  103},
\href{http://arxiv.org/abs/0810.5729}{{\tt arXiv:0810.5729 [hep-th]}}.

\bibitem{Berenstein:1998ij}
D.~E. Berenstein, R.~Corrado, W.~Fischler, and J.~M. Maldacena, ``{The Operator
  product expansion for Wilson loops and surfaces in the large N limit},''
  \href{http://dx.doi.org/10.1103/PhysRevD.59.105023}{{\em Phys.Rev.} {\bf D59}
  (1999)  105023}, \href{http://arxiv.org/abs/hep-th/9809188}{{\tt
  arXiv:hep-th/9809188 [hep-th]}}.

\bibitem{Young:2008ed}
D.~Young, ``{BPS Wilson Loops on S**2 at Higher Loops},''
  \href{http://dx.doi.org/10.1088/1126-6708/2008/05/077}{{\em JHEP} {\bf 0805}
  (2008)  077}, \href{http://arxiv.org/abs/0804.4098}{{\tt arXiv:0804.4098
  [hep-th]}}.

\bibitem{Pestun:2002mr}
V.~Pestun and K.~Zarembo, ``{Comparing strings in AdS(5) x S**5 to planar
  diagrams: An Example},''
  \href{http://dx.doi.org/10.1103/PhysRevD.67.086007}{{\em Phys.Rev.} {\bf D67}
  (2003)  086007},
\href{http://arxiv.org/abs/hep-th/0212296}{{\tt arXiv:hep-th/0212296
  [hep-th]}}.

\end{thebibliography}\endgroup
\end{document}